\documentclass{ceurart}

\usepackage{xurl}

\begin{document}

\copyrightyear{2026}
\copyrightclause{Copyright for this paper by its authors.
Use permitted under Creative Commons License Attribution 4.0
International (CC BY 4.0).}

\conference{ALIT4ALL 2026: 2nd International Workshop on AI Literacy Education For All}

\title{Teaching Students to Question the Machine}
\subtitle{Short-Term Effects of an AI Literacy Workshop on Middle-School Students’ Regulation of LLM Interaction}


 \author[1]{Olivier Clerc}[
 email=olivier.clerc@inria.fr,
 ]
\cormark[1]
\author[2]{Rania Abdelghani}
\author[1]{Chloé Desvaux}
\author[1]{Eliott Poisson}
\author[1]{Pierre-Yves Oudeyer}
\fnmark[1]
\author[1]{Hélène Sauzéon}
\fnmark[1]
\address[1]{INRIA research center University of Bordeaux, Talence, France}
\address[2]{Hector Research Institute, University of Tübingen, Germany}
\cortext[1]{Corresponding author.}
\fntext[1]{Equal supervision.}

\begin{abstract}
Generative artificial intelligence (GenAI) systems are increasingly used by students, making AI literacy an urgent educational goal. In the context of large language models (LLMs), AI literacy involves not only understanding how these systems work and fail, but also applying that understanding to monitor prompts, evaluate responses, and decide when to revise, reject, or continue an interaction. Yet classroom evidence on how to support these applied aspects of AI literacy in middle school remains limited.

We report a quasi-experimental classroom study examining whether a two-hour AI literacy workshop shaped students' regulation of LLM use during science problem-solving tasks. A total of 116 students in grades 8--9 completed six LLM-supported science problems. The intervention group attended a workshop two days earlier that introduced how LLMs work and fail, together with practical guidance on prompting and response evaluation; the control group completed the same tasks without attending the workshop.

Students who attended the workshop showed more adaptive interaction behaviors: they were less likely to accept underspecified prompts, more likely to ask follow-up questions after insufficient responses, and more sensitive to prompt specification when judging response correctness. They also achieved modestly higher final task-performance scores. In contrast, GenAI and metacognitive self-report scores did not robustly predict behavioral regulation or final task performance. These findings suggest that brief AI literacy instruction can shape middle-school students' short-term regulation of LLM-supported science problem solving, while also highlighting the need for longer-term studies of retention, transfer, and learning beyond immediate task performance.
\end{abstract}


\maketitle

\section{Introduction}

Generative artificial intelligence (GenAI) systems are increasingly entering students' everyday learning practices. Recent surveys show rapid uptake of GenAI tools across age groups, including teenagers and students in secondary science, technology, engineering, and mathematics (STEM) contexts~\cite{PewTeens2025,Valeri2025}. Students now use large language model (LLM)-based chatbots for activities such as generating explanations, asking questions, and solving subject-specific problems~\cite{Zhang2024}, creating new opportunities for adaptive support and personalized learning~\cite{Rahman2023,li2024bringing}. At the same time, these opportunities depend on how students engage with model outputs. As LLMs become more fluent, useful, and persuasive, students may be increasingly willing to delegate key reasoning steps to the model rather than generate, evaluate, or revise explanations themselves. For learning, this cognitive offloading may be especially consequential, because it can reduce effortful processing and weaken opportunities to develop critical thinking and self-regulated inquiry~\cite{risko2016cognitive,Gerlich2025,Zhai2024}. The educational value of GenAI therefore depends not only on what these systems generate, but also on how learners regulate their interaction with them~\cite{Wardat2023,adiguzel2023revolutionizing}.

This challenge has made AI literacy an increasingly important educational objective. AI literacy is commonly conceptualized as a set of competencies that includes understanding how AI systems work, recognizing their limitations, using them effectively, and evaluating their social and ethical implications~\cite{chiu2024artificial,yim2024critical}. In LLM-supported problem solving, this requires more than declarative knowledge: learners must translate goals into prompts, evaluate outputs, and decide whether to revise, reject, or continue the exchange. In this paper, we use \emph{regulation of LLM interaction} to refer to this task-level monitoring-and-control process.
This process draws on metacognitive monitoring and control~\cite{tankelevitch2024metacognitive,Yeh2025}, but is specific to the demands of generative AI interaction, where learners must regulate both their own problem-solving process and the system's contributions. Our definition is deliberately narrower than the full AI literacy construct. It does not cover all forms of GenAI regulation, such as detecting sycophantic behavior, deciding when not to use AI at all, or verifying factual accuracy with external sources. These competencies are important, but they were outside the scope of the present intervention and behavioral measures.

Middle-school students may particularly need support in developing this applied dimension of AI literacy. Their critical evaluation and self-regulatory skills are still developing, and persuasive LLM outputs may further encourage over-trust or shallow engagement~\cite{Zhai2024,Bessas2025}. Consistent with this concern, Abdelghani et al.~\cite{abdelghani2025investigating} found that middle-school students often struggled to select effective prompts or challenge unsatisfactory answers when using ChatGPT for science problem solving. Despite having access to an LLM, students produced deep and accurate final explanations in only about half of the tasks on average, suggesting that tool access alone was insufficient. Students with more positive attitudes toward AI were less able to distinguish effective from underspecified prompts. In contrast, self-reported metacognitive regulation and more critical perceptions of AI were associated with behaviors linked to higher-quality final answers, such as better prompt discrimination and more follow-up questioning.

These findings point to a need for AI literacy interventions that are practical, classroom-feasible, and behaviorally assessable. Because students already use LLMs but often struggle to regulate their interactions, instruction should go beyond declarative knowledge. Yet controlled classroom evidence on brief, scalable AI literacy interventions remains limited in K--12 settings~\cite{Guo2025}, and existing school-based studies are often exploratory, small, or lack comparison groups~\cite{Bitzenbauer2023,puppart2025short}.

To address this gap, we conducted a quasi-experimental classroom study in a French middle school ($N=116$, grades 8--9, ages 13--15) to examine whether a two-hour AI literacy workshop shaped students' regulation of LLM use during science problem solving. The workshop combined conceptual content on how LLMs work and fail with practical activities designed to preserve students' agency during interaction.
Importantly, the workshop did not rehearse the experimental science tasks. Instead, students practiced checking whether an initial prompt was aligned with the intended task, whether the LLM response satisfied that goal, and whether they needed to revise the prompt or continue the interaction. Two days later, students completed six science problem-solving tasks with access to a GenAI system; a control group completed the same session without having attended the workshop. In these tasks, suggested prompts were either well-specified or underspecified for the science problem, without students being informed of this manipulation.
The outcome session therefore tested whether students transferred a general regulatory stance---keeping agency over prompting, response evaluation, and follow-up---to the experimental task. This design allowed us to observe whether students accepted or reformulated suggested prompts, whether they asked follow-up questions, and how accurately they judged the correctness of LLM responses.

We treated these behavioral indicators as the primary outcomes of interest. Final answer quality (students' exercise scores) was analyzed as a downstream task-performance measure. We also examined whether GenAI self-reports and a general metacognitive awareness questionnaire predicted task performance. This allowed us to compare behavioral indicators of LLM-interaction regulation with offline self-report measures.

The study was guided by three research questions:
\begin{enumerate}
    \item Does a brief AI literacy workshop change middle-school students' observable regulation of LLM interaction during science problem-solving tasks?
    \item Are students' regulation behaviors associated with their final task performance?
    \item Do GenAI self-reports and general metacognitive awareness predict students' behavioral regulation and final task performance?
\end{enumerate}

\section{Method}

\subsection{Design and participants}

The study was conducted in one French middle school during regular class hours. Because the intervention was implemented within existing school schedules, individual random assignment was not feasible. Students were assigned to the intervention or control condition at the class level according to timetable availability. Both groups were drawn from the same school and were tested during the same period.

A total of 162 students in grades 8 and 9 (typical age range 13--15 years) initially participated. Due to end-of-year scheduling constraints, 27 students who attended the workshop were absent during the test phase and were excluded from the main analyses. An additional 19 students were excluded because they did not complete all six exercises. The final sample comprised 116 students: 40 in the control group (24 grade 8, 16 grade 9) and 76 in the intervention group (50 grade 8, 26 grade 9). The unequal group sizes reflect the use of intact classes and differences in attendance across the workshop and test sessions.

Within the intervention group, two students provided inconsistent identifiers between the pre- and post-intervention GenAI questionnaires. These students were retained for the main behavioral and task-performance analyses, but excluded from pre--post questionnaire analyses. Accordingly, questionnaire analyses were conducted on $n=40$ control students and $n=74$ intervention students.

\subsection{Procedure overview}

Control students completed a 90-min exercise session individually, solving science problems with access to an LLM as a support tool. Intervention students completed the same exercise session, preceded two days earlier by a two-hour AI literacy workshop. A timeline and classroom procedure are provided in the Supplementary Material (SM Figs.~S1--S2).

Before the exercise session, all students completed self-report questionnaires measuring GenAI perceptions and metacognitive awareness. During the exercise session, students completed six science problem-solving tasks. For each task, they decided whether to use a suggested prompt, interacted with the LLM, evaluated the response, optionally asked a follow-up question, and wrote a final answer in their own words.

\subsubsection{AI literacy workshop}

The intervention was a two-hour, classroom-feasible AI literacy workshop designed to strengthen both conceptual understanding of LLM-based GenAI systems and applied regulation of LLM interaction. The conceptual component introduced age-appropriate explanations of how LLMs generate text, why they can produce incomplete or incorrect answers, and why fluent responses should not be accepted uncritically.
The practical component did not rehearse the later experimental science tasks. Instead, it focused on preserving students' agency during interaction with an LLM. Students first completed a collective response-evaluation activity in which they discussed common problems in LLM outputs.
Students then worked in small groups on prompting activities from domains distinct from the experimental science tasks. For each activity, they received a task goal and an initial prompt, predicted whether the prompt was likely to elicit a satisfactory response, queried the LLM, evaluated the answer, and revised the prompt or continued the interaction when needed. These activities trained students to maintain agency during their LLM interaction. Examples of workshop materials are provided in the Supplementary Material (SM S3), and the full workshop slide deck and facilitation script are available in the accompanying OSF repository.

\subsubsection{Science problem-solving tasks and LLM environment}

\begin{figure}[t]
  \centering
  \includegraphics[width=\linewidth]{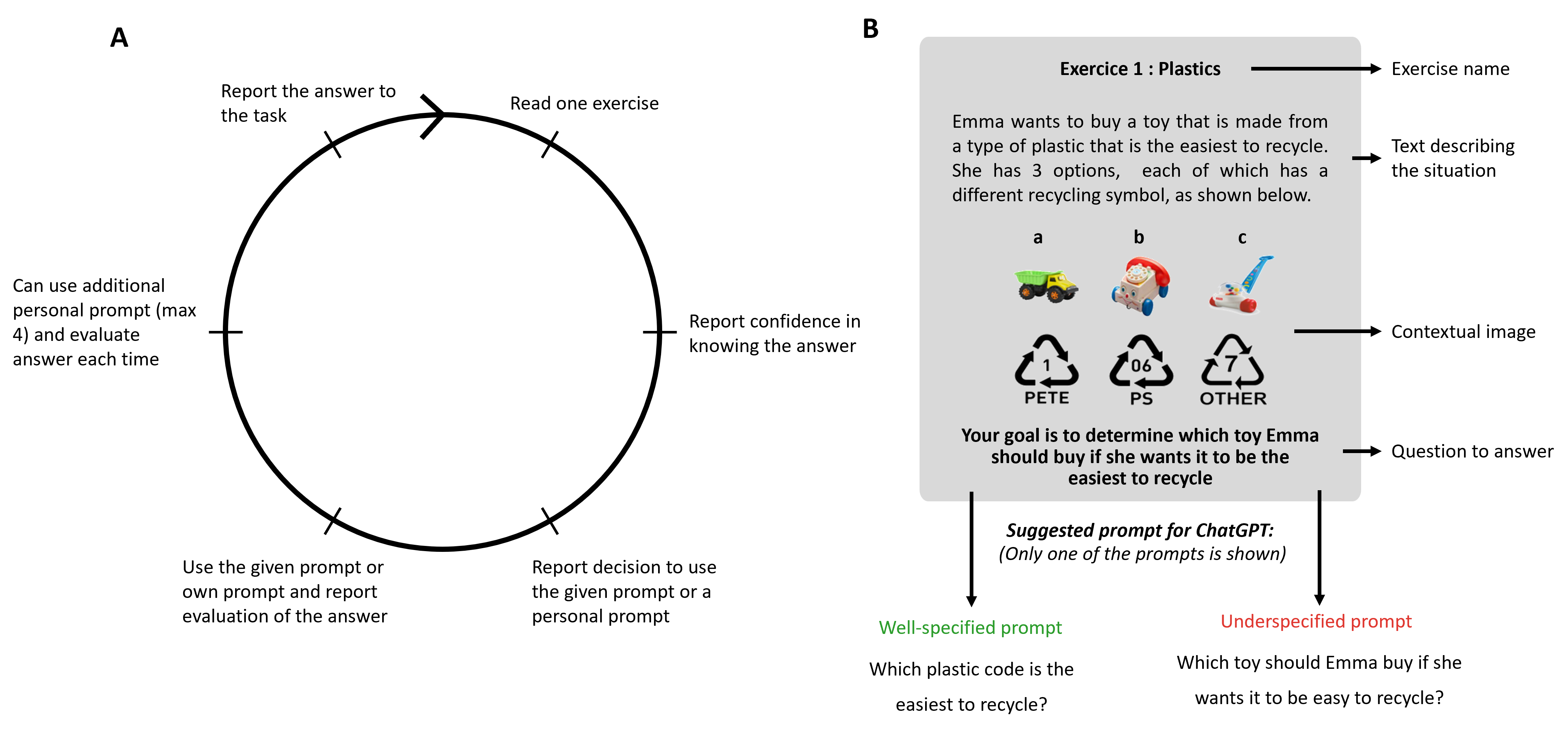}
  \caption{\textbf{Student--LLM workflow and exercise structure.} \textbf{(A)} For each exercise, students rated confidence (3-point), decided whether to use the suggested prompt, iteratively queried and evaluated the LLM (3-point; optional follow-up), and wrote a final answer; repeated across six exercises. \textbf{(B)} Example sheet and prompt manipulation. The suggested prompt was either well-specified or underspecified (one shown per exercise).}
  \label{fig:task}
\end{figure}

The exercise session was based on the science problem-solving tasks described in Abdelghani et al.~\cite{abdelghani2025investigating}. Each student completed six science exercises within 90 minutes (Fig.~\ref{fig:task}). The six exercises were randomly selected from a pool of twelve tasks and presented in random order to limit collaborative solving between students.

Each exercise included a single suggested prompt. Suggested prompts were either well-specified, meaning that they contained the key contextual elements needed for the exercise, or underspecified, meaning that they omitted important information and were likely to elicit generic or incomplete responses. Prompt specification was pseudorandomized such that each student encountered three well-specified and three underspecified prompts. Students were not informed about this manipulation. Exercises were provided on paper to prevent direct copy-paste of the full problem statement into the chatbot interface.

Students worked individually at a computer with two browser tabs open: an LLM-based chatbot interface and a web questionnaire guiding the procedure. For each exercise, students could interact iteratively with the LLM and evaluate its responses before producing a final written answer in their own words.

Because access constraints sometimes required students to use different public interfaces, the LLM environment was not perfectly homogeneous across task attempts. They primarily used GPT-4o mini \textit{via} DuckDuckAI; when this interface failed, some used other public chatbot interfaces running GPT-4o, GPT-4.1-nano, or GPT-3.5-turbo (robustness checks and interface use: SM S4.2–S4.4). To address this, we conducted robustness checks on the prompt manipulation itself, verifying that prompts designed as underspecified still elicited low-quality responses and well-specified prompts high-quality responses independently of the possible models used by the students. We report interface use in SM S4.4, Table S5.

\subsection{Measures}

\subsubsection{Behavioral indicators of LLM-interaction regulation}

We operationalized regulation of LLM interaction through three behavioral indicators collected during each exercise. First, \emph{query-level control} was measured through prompt acceptance: students decided whether to use the suggested prompt or replace it with their own formulation. Rejecting an underspecified prompt was interpreted as adaptive regulation, whereas accepting it indicated lower query-level control. Second, \emph{response evaluation} was measured through students' correctness judgments after receiving the LLM answer. Third, \emph{interaction control} was measured through follow-up questioning: students could ask additional questions after the initial LLM response, a behavior expected to be especially useful when the initial response was insufficient. These measures were treated as behavioral indicators of the applied regulatory dimension of AI literacy, not as measures of the full AI literacy construct.

\subsubsection{Final answer quality}

Students' final written answers were scored on a 0--2 rubric: 0 = incorrect, 1 = partially correct, and 2 = correct. Exercise-level scores were used in analyses linking regulation behaviors to answer quality. Student-level final task performance was computed by summing the six exercise scores and converting the total to a 0--20 scale.

\subsubsection{Self-reported measures}

All students completed two questionnaires prior to the exercise phase. The first was a GenAI questionnaire adapted from Bernabei et al.~\cite{bernabei2023students} following Abdelghani et al.~\cite{abdelghani2025investigating}. It covered six dimensions: Attitude, Trust, Social influence, Fairness \& ethics, Usefulness, and Effort \& ease of use (24 items; 4-point Likert scale; maximum score = 96). It captured students' familiarity with ChatGPT, perceived reliability of its answers, and perceived usefulness for school tasks; for example, items asked whether students knew the limits of ChatGPT and whether they considered its answers reliable. Because this adapted version has not been formally validated for this age group, GenAI self-reports were treated as secondary and interpreted cautiously.

The second questionnaire was the Jr. MAI~\cite{kim2017establishing}, a middle-school measure of metacognitive knowledge and regulation of cognition (18 items; 5-point Likert scale; maximum score = 90). It was included as a broad measure of general metacognitive awareness, not as an LLM-specific regulation instrument. Full item wording and scoring details are provided in SM~S2.

Questionnaires served two purposes. First, GenAI and metacognitive self-report scores were used as individual-difference predictors to test whether students' self-perceptions were associated with behavioral regulation and final task performance. Second, in the intervention group only, GenAI self-reports were collected both before and after the workshop as a secondary outcome to assess whether the workshop changed these self-reported dimensions.

\subsection{LLM-as-judge validation and final-answer scoring}

We used GPT-4o as an LLM-as-judge to validate the prompt manipulation and to score students' final written answers. In both cases, we first evaluated the annotation setup on a random $\approx 10\%$ subset independently rated by human annotators, and applied GPT-4o to the full dataset only after meeting predefined reliability and non-inferiority criteria. Agreement was assessed with Krippendorff's $\alpha$, appropriate for multi-rater ordinal ratings, and with the advantage-probability test of Calderon et al.~\cite{calderon2025alternative}, which compares the model against held-out human annotators.

For prompt-response annotation, agreement was higher when GPT-4o was included ($\alpha=.825$) than for human-only annotations ($\alpha=.763$), and GPT-4o outperformed each held-out human annotator ($\omega=1.00$). For student-answer scoring, GPT-4o was non-inferior to human annotators ($\alpha=.752$ with GPT-4o included vs. $\alpha=.740$ human-only; $\omega=1.00$). Full validation details are reported in SM S4.

\subsection{Data analysis}

Analyses were conducted in Python with two-sided tests and $\alpha=.05$. For repeated task-level observations, generalized linear models used cluster-robust standard errors at the student level. We report odds ratios (ORs) for logistic models, unstandardized coefficients for linear models, and effect sizes where appropriate. False-discovery-rate (FDR) correction was applied to families of questionnaire follow-up tests. Full details are reported in the Supplementary Material, and the anonymized analysis notebook used to generate the results and figures is provided in the accompanying OSF repository.

We first tested whether the intervention affected the three behavioral indicators of LLM-interaction regulation. Query-level control was analyzed with a logistic GLM predicting prompt acceptance from condition, prompt specification, their interaction, and pre-query confidence. Response evaluation was analyzed among accepted-prompt trials with a logistic GLM predicting correctness judgments from condition, prompt specification, and their interaction. Interaction control was analyzed among accepted underspecified-prompt trials by comparing student-level follow-up-question rates between groups.

We then tested whether regulation behaviors were associated with final answer quality. For each prompt type, we computed within-student reject-minus-accept contrasts in answer quality and compared these contrasts between groups. We also tested whether follow-up-question rates after accepted underspecified prompts predicted mean answer quality differently by condition.

Because final task scores were bounded and non-normal in the intervention group, the main group comparison used a Mann--Whitney test. We also computed signal-detection indices ($d'$ and response bias $\beta$) to quantify students' selectivity in accepting well-specified rather than underspecified prompts, and tested whether these indices predicted final task performance. Finally, we examined whether GenAI and metacognitive self-reports predicted behavioral regulation or final task performance.

\subsection{Ethics and data protection}

The study was approved by an institutional ethics committee; identifying details are withheld for double-blind review. Parent/guardian consent and student assent were obtained before participation. No personally identifying data were collected. Students did not use personal accounts to access the LLM interface and were instructed not to enter personal information into the chatbot.

To preserve students' privacy, we did not collect or access the full transcripts of their interactions with the LLM. The dataset only contained students' responses to the study questionnaire, including their prompt-use decisions, correctness judgments, follow-up-question indicators, confidence ratings, and final written answers. To preserve anonymity and comply with feasibility constraints in the school context, we did not collect gender, race/ethnicity, or other demographic identifiers beyond grade level. As a result, the dataset could not be used to test whether intervention effects differed across demographic subgroups.

\section{Results}

\subsection{The intervention changed students' regulation of LLM interaction}

We first examined whether the workshop affected the three behavioral indicators of LLM-interaction regulation.

\emph{Query-level control:} A logistic GLM predicting prompt acceptance showed that intervention students were less likely than control students to accept suggested prompts overall (OR $=0.47$, $p=.022$; average marginal effect $=-.155$, $p=.018$; Fig.~\ref{fig:workshop}C). Well-specified prompts were more likely to be accepted than underspecified prompts (OR $=1.85$, $p=.040$). The condition $\times$ prompt-specification interaction was not significant ($p=.228$), but follow-up contrasts showed that the group difference was concentrated on underspecified prompts: intervention students accepted these prompts less often than controls (51.5\% vs. 66.7\%; OR $=0.47$, $p=.022$), whereas groups did not differ reliably for well-specified prompts (74.4\% vs. 79.2\%; $p=.385$).

\emph{Response evaluation:} Students in both groups were more accurate at judging the correctness of AI responses when prompts were well-specified rather than underspecified (OR $=2.06$, $p=.001$), but this effect was larger in the intervention group (interaction OR $=2.52$, $p=.012$; Fig.~\ref{fig:triptic2}B), indicating that intervention students' evaluation judgments were more accurate. 

\emph{Interaction control:} Among accepted underspecified-prompt trials, intervention students asked follow-up questions more often than controls (59.2\% vs. 27.9\%; Welch $t=3.80$, $p<.001$, $d=0.80$; Fig.~\ref{fig:triptic2}C).

Additional analyses are provided in SM S5.3--S5.5.

\subsection{Regulation behaviors were associated with final answer quality}

\begin{figure}[t]
  \centering
  \includegraphics[width=\linewidth]{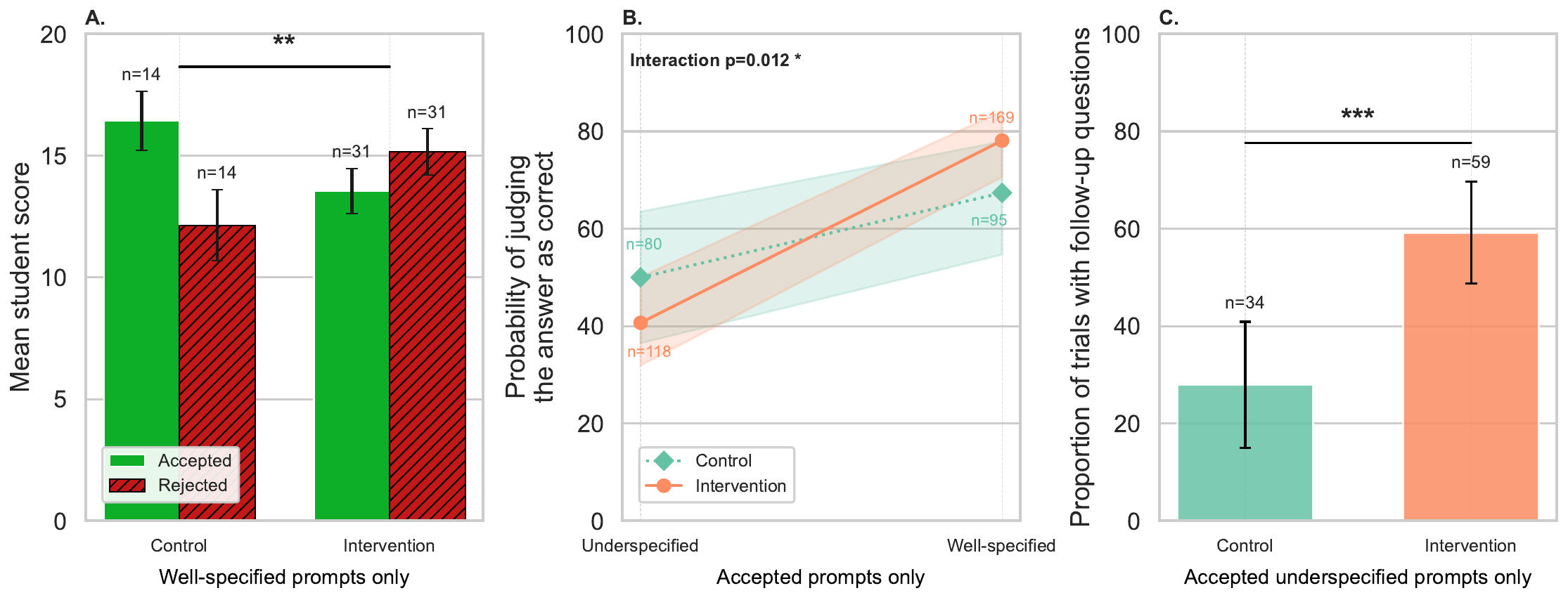}
  \caption{\textbf{Effects of the intervention on prompt replacement, follow-up behavior, and correctness judgments.} \textbf{(A)} Within-student performance by accept vs. reject on well-specified prompts. Rejecting well-specified prompts was more costly in the control group. \textbf{(B)} Probability of judging an AI answer as correct (GLM; 95\% CI) by prompt specification. Trained students made more accurate correctness judgments. \textbf{(C)} Follow-up questioning after accepted underspecified prompts by group. Trained students asked more follow-up questions after underspecified prompts.}
  \label{fig:triptic2}
\end{figure}

We next examined whether students' regulatory behaviors were associated with final answer quality. For well-specified prompts, rejecting rather than accepting the suggested prompt was costly in the control group but not in the intervention group. The reject-minus-accept contrast differed significantly between groups ($M=-4.29$ in control vs. $M=1.61$ in intervention; Welch $t=2.83$, $p=.009$, $d=0.91$; Fig.~\ref{fig:triptic2}A). This suggests that control students often replaced well-specified prompts with less effective alternatives, whereas intervention students could reformulate prompts without a comparable performance penalty.

For underspecified prompts, rejected trials scored higher than accepted trials descriptively in both groups, but the between-group difference in this contrast was not significant ($p=.194$). Because follow-up questions are especially relevant when students initially accept an underspecified prompt, we examined this subset directly. Among students who accepted underspecified prompts, follow-up-question rate was more strongly associated with final answer quality in the intervention group than in the control group ($b=6.43$, 95\% CI [0.36, 12.50], $p=.038$). Within the intervention group, this behavior was associated with higher-quality answers (approximately $+5$ points out of 20), whereas no such benefit was observed in the control group.

\subsection{Final task performance showed a modest group difference}

\begin{figure}[t]
  \centering
  \includegraphics[width=\linewidth]{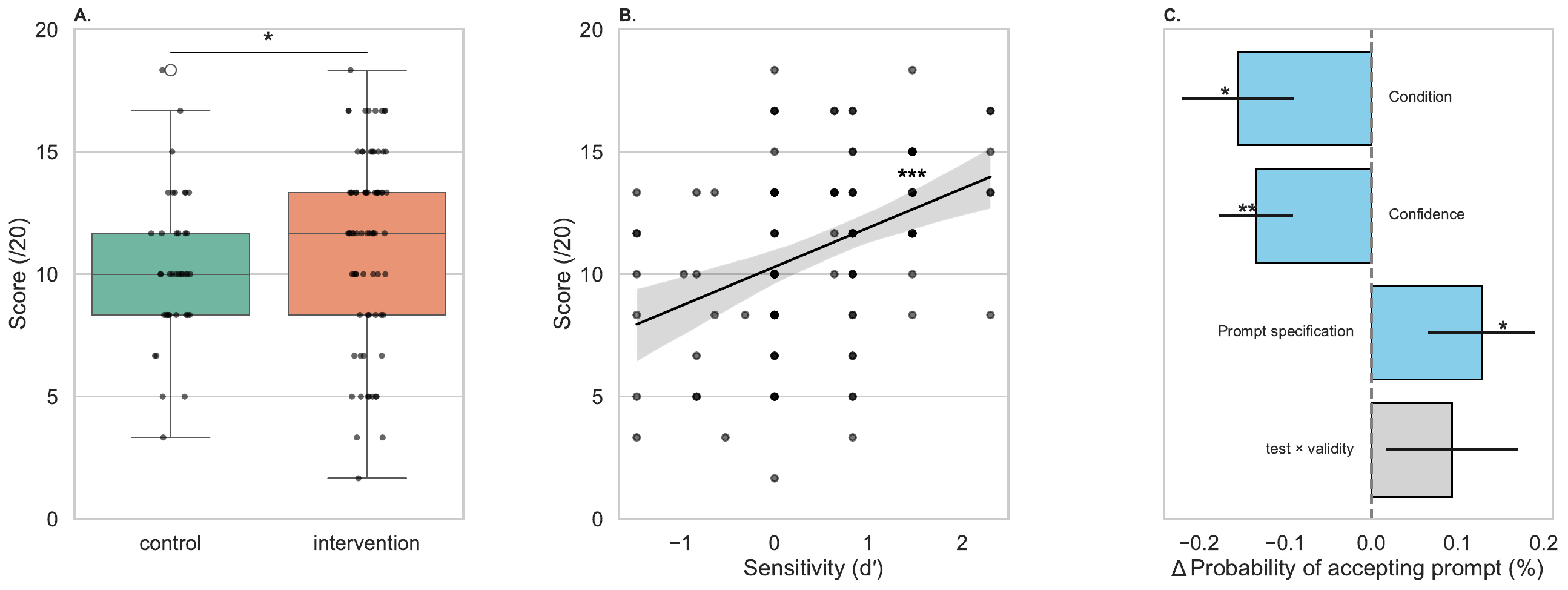}
  \caption{\textbf{Workshop effects on performance and prompt acceptance.} \textbf{(A)} Final-answer performance (0--20) by group. Intervention students achieved higher final-answer scores. \textbf{(B)} Relationship between performance and prompt-discrimination sensitivity ($d'$). Performance increased with prompt discrimination. \textbf{(C)} Predictors of accepting the suggested prompt (GLM).}
  \label{fig:workshop}
\end{figure}

Final task-performance scores differed between groups, with higher scores in the intervention group than in the control group ($M=11.38$, $SD=3.87$, $n=76$ vs. $M=10.29$, $SD=3.04$, $n=40$; Mann--Whitney $U=1871$, $p=.040$, $r_z=.19$, rank-biserial $r=.23$; Fig.~\ref{fig:workshop}A). This effect was modest, and absolute performance remained limited in both groups despite access to a GenAI system. However, maximum scores were high (18.3/20), indicating that the tasks were feasible under the scoring and time constraints.

To examine whether final answer quality was related to students' selectivity in accepting well-specified versus underspecified prompts, we computed signal-detection indices. Sensitivity ($d'$) was numerically higher in the intervention group than in the control group, but this difference was not significant ($M=0.53$ vs. $0.28$; Welch $t=1.45$, $p=.150$, $d=0.27$). Response bias ($\beta$) did not differ between groups ($p=.746$). Across students, higher sensitivity predicted higher final task-performance scores ($b=1.76$, 95\% CI [1.06, 2.46], $p<.001$), as did a more conservative response bias ($b=1.65$, 95\% CI [0.25, 3.05], $p=.021$). Once $d'$ and $\beta$ were included in the model, the direct condition effect was no longer significant ($p=.279$). The condition $\times d'$ interaction was positive but did not reach significance ($b=1.36$, $p=.072$). Exploratory within-group models showed that both $d'$ and $\beta$ predicted final task performance in the intervention group, whereas neither predictor was reliable in the control group. These results suggest that the workshop may have helped students act on prompt-discrimination cues, rather than simply improving their ability to detect underspecified prompts. In other words, the intervention may have strengthened the link between noticing a problem and selecting an appropriate regulatory action, such as reformulating the prompt or asking a follow-up question.

\subsection{GenAI self-report scores changed only weakly}

Baseline GenAI self-report dimensions did not differ reliably between groups. A one-way MANOVA on pre-test dimensions was not significant (\emph{Pillai's trace} = .083, $F(6,107)=1.61$, $p=.151$), and FDR-corrected univariate comparisons showed no significant baseline group differences (all $p_{\mathrm{FDR}} \geq .189$).

Within the intervention group, a multivariate test of pre--post change scores showed a small overall shift in GenAI self-reports (\emph{Pillai's trace} = .195, $F(6,68)=2.75$, $p=.019$). However, no individual dimension survived FDR correction. Attitude and Fairness \& ethics showed small uncorrected increases (Attitude: $M_{\Delta}=0.11$, $p=.018$, $p_{\mathrm{FDR}}=.054$; Fairness \& ethics: $M_{\Delta}=0.13$, $p=.011$, $p_{\mathrm{FDR}}=.054$), but these effects should be interpreted cautiously. Thus, the workshop's effects were clearer for behavioral indicators of LLM-interaction regulation than for self-reported GenAI perceptions.

\subsection{Self-report measures did not robustly predict behavioral regulation or final task performance}

We first tested whether GenAI and metacognitive self-report scores predicted the student-level behavioral indicators of LLM-interaction regulation described above. None of the associations survived FDR correction (all $p_{\mathrm{FDR}} \geq .580$). Within the intervention group, changes in the six GenAI self-report dimensions explained little variance in prompt-discrimination sensitivity ($R^2=.11$), and dimension-level predictors were not robust after correction.

Self-report scores also did not explain final task performance. Across students, GenAI self-report scores were not correlated with final task-performance scores ($r=.01$, $p=.881$, $R^2<.001$, $n=116$). Metacognitive self-report scores were likewise unrelated to final task-performance scores ($r=.04$, $p=.649$, $R^2=.002$, $n=116$). Together with the behavioral analyses above, these results suggest that observable regulation during LLM use was more informative than offline self-reports for explaining students' final answer quality.

\section{Discussion}

This study examined whether a brief AI literacy workshop could shape middle-school students' regulation of LLM use during science problem solving. The main finding is that the workshop changed students' observable interaction behaviors. Compared with control students, intervention students were less likely to accept underspecified prompts, their response-evaluation judgments were more sensitive to prompt specification, and they asked more follow-up questions when initial prompts were underspecified. These behavioral changes were accompanied by a modest downstream difference in final task performance, but absolute scores remained limited in both groups. Thus, the study should not be interpreted as showing durable learning effects. Rather, it provides classroom-based evidence that short AI literacy instruction can affect how students regulate LLM-supported problem solving under realistic school constraints.

\subsection{AI literacy as regulatory action}

The workshop was designed to support an applied dimension of AI literacy: maintaining agency while interacting with an LLM. The results suggest that this regulatory stance transferred to the science problem-solving tasks. Intervention students were more selective with underspecified prompts, their correctness judgments were more sensitive to prompt specification, and they asked more follow-up questions when the initial prompt was likely to produce an insufficient answer.

These effects appear to reflect changes in regulatory action more than large improvements in prompt-discrimination ability. Although intervention students accepted underspecified prompts less often, their overall sensitivity to prompt specification was only numerically higher than controls'. In contrast, prompt-discrimination sensitivity and response bias predicted final task performance more reliably in the intervention group. This suggests that the workshop may have strengthened the link between noticing a problem and acting on it. The fact that control students suffered a performance cost when replacing well-specified prompts further suggests that unguided prompt revision can be harmful when students do not know what information needs to be preserved.

\subsection{Task performance is improved but is not evidence of durable learning}

Intervention students achieved modestly higher final task-performance scores than control students, but this outcome should be interpreted carefully. Students completed the tasks with access to an LLM, and the test occurred only two days after the workshop. The study therefore shows improved regulation during LLM-supported problem solving, not durable learning gains.

The modest absolute performance in both groups also reinforces this point. Even with access to GenAI, students' average final scores remained around 11--12 out of 20. This suggests that access to an LLM alone is not sufficient for producing high-quality answers. Students still need to identify what information is relevant, evaluate whether the model's response is adequate, and synthesize a final answer in their own words. These demands are challenging for middle-school students~\cite{winograd1984strategic} and may be especially important when GenAI outputs appear fluent and authoritative. Because the activity was low-stakes (not graded), reduced motivation may also have contributed~\cite{demars2000test} to this; importantly, these constraints applied to both groups and therefore do not undermine the between-group effect.

\subsection{Self-reports were weak indicators of applied AI literacy}

GenAI and metacognitive self-report scores did not predict students' behavioral regulation or final task performance. It suggests that offline self-reports may capture students' perceived competence and not actual regulation during interaction. This interpretation is consistent with recent evidence that self-report measures of metacognition often correlate weakly with online monitoring judgments during task performance, whereas behavioral indicators of monitoring and control can be more informative for explaining learning-related outcomes~\cite{seban2025relationship}. It also aligns with broader evidence that self-assessments can overestimate competence~\cite{dunning2004flawed}, including in the domain of digital skills~\cite{Palczyska2021}.

The workshop also produced only limited changes in GenAI self-reports. Although the multivariate pre--post test suggested a small overall shift, no individual dimension survived correction for multiple comparisons. In contrast, clearer effects were observed in students' behavior during the LLM-supported task. This pattern suggests that changes in applied AI literacy may not be well captured by short-term changes in self-reported attitudes, confidence, or perceived knowledge. 

At the same time, the null self-report findings should be interpreted cautiously. The GenAI questionnaire was adapted from prior work and has not been formally validated for this age group. The Jr. MAI measures general metacognitive awareness, not LLM-specific regulation. Future work should develop and validate AI-literacy instruments that distinguish conceptual knowledge, attitudes, ethical awareness, perceived competence, and applied regulation during interaction.

\subsection{Implications for classroom AI literacy education}

The study has three implications for AI literacy education in middle school. First, it suggests that brief, classroom-feasible interventions can change students' LLM interaction behaviors, at least in the short term. This extends Abdelghani et al.~\cite{abdelghani2025investigating}, who showed that middle-school students often struggle to select effective prompts and challenge unsatisfactory answers when using ChatGPT for science problem solving. The control group in the present study showed similar difficulties, consistent with evidence that non-experts often struggle to design robust prompts without structured support~\cite{zamfirescu2023johnny}. A two-hour workshop might not be expected to produce stable expertise, but it can introduce regulatory habits that students may begin to apply during LLM-supported tasks.

Second, AI literacy interventions should target not only conceptual knowledge about how AI systems work, but also practical regulation skills and learner agency. This matters because students may possess general questioning and explanation-evaluation skills in non-GenAI contexts~\cite{mills2011determining,mills2019want}, but may not spontaneously apply them when confronted with fluent and confident-looking model outputs~\cite{Zhai2024}. In this sense, the goal of AI literacy education is not simply to teach students about AI, but to help them remain active agents during AI-supported work.

Third, AI literacy interventions should be evaluated with behavioral measures in addition to self-reports. The present findings suggest that self-reports may miss important differences in actual use. This is especially important if AI literacy is understood as a competence that includes procedural knowledge, not only declarative knowledge.

\subsection{Limitations and future directions}

Several limitations should be noted. First, the study was quasi-experimental rather than individually randomized. Groups came from the same school, were tested during the same period, and did not differ on pre-test GenAI self-reports, but students were assigned by class schedules and group sizes were unequal. Unmeasured baseline differences, including science ability, therefore cannot be ruled out.

Second, the study measured short-term effects only, as the test occurred two days after the workshop. Future work should examine retention, repeated practice, transfer to new tasks, and performance without LLM support. Third, the study was conducted in one school and focused on science problem solving, so generalization to other subjects, schools, and student subgroups remains to be tested.

Fourth, the prompt-acceptance task does not fully mirror ecological student uses of LLMs. In many classroom contexts, students would write their own prompts from scratch rather than decide whether to accept a suggested prompt. Prompt acceptance should therefore be interpreted as an experimental probe of whether students noticed missing task information. By contrast, evaluating the adequacy of the model's response and deciding whether to ask a follow-up question are likely to remain central forms of regulation whenever students use LLMs for learning.

Fifth, the LLM environment was not fully homogeneous because students used different public interfaces when access constraints required it. However, this heterogeneity reflects ecological conditions, and our prompt-validation procedure confirmed that the experimental manipulation was robust across models. Finally, our behavioral measures captured broad regulatory decisions but not the linguistic content of prompts or follow-up questions. Future work could combine these indicators with privacy-preserving transcript analysis.

Overall, the findings suggest that brief, classroom-feasible AI literacy instruction can support students' applied regulation of LLM interaction, but longer-term and broader studies are needed to determine whether such regulation leads to durable learning benefits.

\section*{Supplementary Materials, Data, Code, and Workshop Materials}

Supplementary materials, workshop materials, analysis code, and anonymized data are available in an OSF repository: \url{https://osf.io/fyu7c/overview}. The repository is organized into three folders: \texttt{supplementary\_materials}, \texttt{workshop\_materials}, and \texttt{data\_and\_code}. The \texttt{workshop\_materials} folder includes the workshop slide deck and facilitation script, currently available in French. The \texttt{data\_and\_code} folder includes the anonymized dataset and the code used to reproduce the analyses and figures reported in the paper.

\section*{Declaration on Generative AI}
During the preparation of this work, the author(s) used generative AI tools for language editing and formatting assistance. After using these tools, the author(s) reviewed and edited the content as needed and take full responsibility for the publication's content.

\bibliography{references}

\end{document}